\documentclass[
 aip,
 jap,%
 amsmath,amssymb,
 reprint,%
]{revtex4-1}
\usepackage[final]{changes}
\usepackage{graphicx}
\usepackage{chngcntr}
\counterwithout{figure}{section}
\usepackage{dcolumn}
\usepackage{bm}
\usepackage{circuitikz}
\usepackage{tikz}

\usepackage{mathtools}
\DeclarePairedDelimiter\abs{\lvert}{\rvert}
\usepackage{comment}

\begin{document}


\title{Wireless Power Transfer via Dielectric Loaded Multi-moded Split Cavity Resonator}

\author{Sameh. Y. Elnaggar}
\email{samehelnaggar@gmail.com}
\affiliation{Department of Electrical and Computer Engineering, Royal Military College of Canada, Kingston, ON, Canada.}
\author{Chinmoy Saha}
\email{csaha@ieee.org}
\affiliation{Department of Electrical and Computer Engineering, Royal Military College of Canada, Kingston, ON, Canada.}
\affiliation{Indian Institute of Space Science and Technology,Thiruvananthapuram, Kerala, India.}
\author{Yahia Antar}
\email{antar-y@rmc.ca}
\affiliation{Department of Electrical and Computer Engineering, Royal Military College of Canada, Kingston, ON, Canada.}

\date{\today}

\begin{abstract}
Wireless power transfer via a dielectric loaded multi-moded split cavity resonator (SCR) is proposed in this article. Unlike conventional inductive resonant coupling, the scheme enables the control of both the real and imaginary parts of the transfer impedance. It is demonstrated through measurements, analytical models and extensive full-wave simulation that the inclusion of dielectric resonators (DRs) tuned to the SCR $TE_{012}$ mode, significantly enhances the system figure of merit, optimal efficiency and maximum power transferred to the load. The effect of the DRs is shown to be related to the resonant coupling of the DRs $TE_{01\delta}$ and SCR modes, resulting in an electromagnetic induced transparency-like window. An efficiency of 70\% is achieved when the transfer distance is 7 cm, or half wavelength. Additionally, it was shown that the efficiency is above 40\% over a relatively wide bandwidth and a wide range of optimum load impedance. A circuit model is developed that enables the decomposition of the two port network parameters into their modal contributions. Hence it allows the comparison with conventional inductive resonant coupling systems on the fundamental level. Additionally, a Vector Fitting based method is proposed to calculate the circuit parameters from the measured scattering parameters.
\end{abstract}

\keywords{}
\maketitle

\section{Introduction}
\label{sec:introduction}
Wireless Power Transfer (WPT) has gained immense interest in the last decade due to its vast potential applications in existing and emerging domains such as in powering appliances, gadgets, electric vehicles and internet of things. Resonant inductive (or capacitive) coupling provide a promising mean of WPT due to its high efficiency over mid-range distances\cite{Kurs2007,Karalis2008}.

Resonant coupling methods rely on the mutual coupling between relatively high $Q$ resonators. The strength of coupling can be quantified via the unit-less coupling coefficient $\kappa$. In terms of circuit elements, $\kappa$ is directly proportional to the mutual coupling $M$ between the different resonators. For a general electromagnetic resonators, it is the net overlap of electromagnetic fields of resonant modes\cite{Elnaggar2015JAP}. To reduce unnecessary intrinsic losses, it is desirable to maximize the source and receiver intrinsic $Q$ factors. Due to their very high $Q$ values, the modes of Dielectric Resonators (DRs) have been exploited to transfer power efficiently over short and mid-range distances\cite{Song2016Collosal,Song2016,songsmarttable}. The performance of resonant coupled schemes correlates directly with the product $\kappa Q$, which represents the system Figure of Merit (FOM).

On the other hand, the modes of cavity resonators were exploited to transfer power within an enclosure \cite{SampleCavity2015,chabalko_apl,mei2016optimal}. This approach allows power to be transferred to almost any point in a 3D region. By capacitively loading a rectangular cavity, quasi-static modes can be supported, where the electric field is localized inside the capacitors banks and, therefore, health hazards are minimized\cite{Chabalko2017}.

A moderately high $\epsilon_r$ DR placed inside a conducting cavity was shown to strongly interact with the cavity modes\cite{Elnaggar2014coupled}. Placing two DRs that are practically uncoupled from one another, creates an electromagnetic induced transparency like (EIT-like) window. Through this window, it was theoretically shown that efficient power transmission can be made possible\cite{Elnaggar2017JAP}. Additionally, a split cavity resonator (SCR) was used instead of the fully cavity and it was shown that, via the EIT-like pathway, high efficient WPT is still feasible.

In the current manuscript, we experimentally demonstrate that an SCR loaded with two DRs (DR2SCR) provides efficient WPT. The SCR and DRs dimensions are chosen such that the DRs $TE_{01\delta}$ and SCR $TE_{012}$ modes have the same resonant frequency. Depending on the coupling of the input power to the SCR in the vicinity of the $TE_{012}$ resonance, several modes can be excited, resulting in a multimoded structure. Therefore, the structure can be regarded as a platform that enables the exploration of the contribution of multiple modes, or pathways, to the delivered power. Hence to describe the system behaviour and interpret the experimental results, a framework based on a combination of electromagnetic analysis and network theory is applied. Particularly, network theory allows the decomposition of the terminal parameters into their modal constituents. Additionally as a first order approximation, Coupled Mode Theory (CMT) and circuit modelling are applied to describe the interaction between the SCR $TE_{012}$ and DRs $TE_{01\delta}$ modes\cite{Elnaggar2017JAP}.

Section \ref{sec:Theory} describes the network theory approach to WPT systems and highlights the relevant electrical parameters. It will be shown that the framework permits the abstraction of the FOM and expresses it in terms of generic circuit parameters that are valid for arbitrary WPT systems. Furthermore, the electromagnetic modes are represented by a modal equivalent circuit and the procedure for calculating the ciruit parameters are outlined in detail. The geometry and dimensions of the SCR and DR2SCR structures are presented in Section \ref{sec:scrdr2scr}. Backed by full-wave simulations, the spatial disributions of the EM modes are computed for the SCR and DR2SCR schemes. Section \ref{sec:Measurements} presents experimental results, compares to simulations, discusses the findings, and relates the behaviour to the circuit parameters and interactions of modes. 
\section{Theoretical Background}
\label{sec:Theory}
 As mentioned in Section \ref{sec:introduction}, coupled resonators enable WPT via the possible electromagnetic coupling between them and their high $Q$ values. The FOM of a WPT system can be expressed as the product $\kappa Q$. From a circuit theory perspective, coupled resonators can be modelled by a two port network, thus allowing the FOM to be written in terms of circuit parameters as follows,
 \begin{equation}
 \label{eq:kQ}
 \textnormal{FOM}=\frac{|Z_{21}|}{\sqrt{r_{11}r_{22}-r_{21}^2}}=\sqrt{\frac{r_{21}^2/r_{11}r_{22}+x_{21}^2/r_{11}r_{22}}{1-r_{21}^2/r_{11}r_{22}}},
\end{equation}  
where $Z_{mn}=r_{mn}+jx_{mn}$ is the $(m,n)$ element of the $\mathbf{Z}$ matrix\cite{dionigi2015rigorous,minnaert2017single}. The expression of FOM in (\ref{eq:kQ}) is an extended $\kappa Q$ expression that is valid for an arbitrary two port network. Indeed given a two port network, its maximum efficiency $\eta_\textnormal{max}$ is fully determined via $\textnormal{FOM}$. As was previously shown, if $\textnormal{FOM}$ is represented by the tangent of an angle $2\theta$ (i.e, $\textnormal{FOM}\triangleq\tan 2\theta$), $\eta_\textnormal{max}$ is precisely $\tan^2\theta$\cite{ohira2014extended}. Equation (\ref{eq:kQ}) reveals that the FOM depends on the relative magnitudes of the transfer components $r_{21}$ and $x_{21}$ with respect to the product of self resistor values $r_{11}$ and $r_{22}$. Accordingly, (\ref{eq:kQ}) can be conveniently re-written as,

\begin{equation}
\label{eq:kQnorm}
\textnormal{FOM}=\sqrt{\frac{r_n^2+x_n^2}{1-r_n^2}},
\end{equation}
where $r_n\triangleq r_{21}/\sqrt{r_{11}r_{22}}$ and $x_n\triangleq |x_{21}|/\sqrt{r_{11}r_{22}}$. Therefore, the FOM can be enhanced by increasing $r_n$ and/or $x_n$.
\begin{figure*}[!htb]
\center
\includegraphics[width=2\columnwidth]{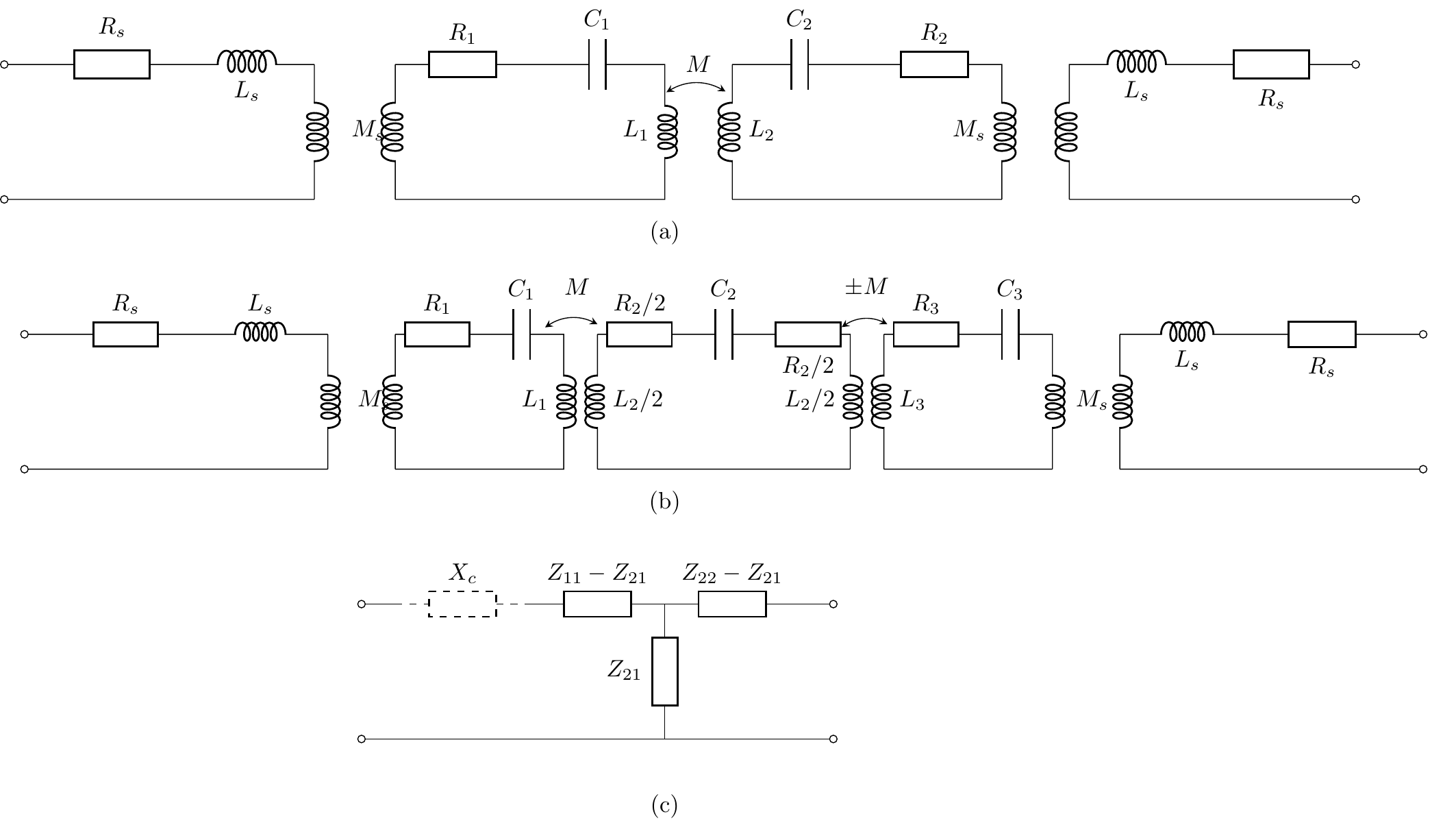}
\label{fig:circuittwport}
\caption{(a) A circuit model of inductively coupled resonators used in WPT systems. (b) A three inductively coupled resonators. (c) A T representation of a generic two port network with the compensation reactance $X_c$ connected.}
\end{figure*}

 The circuit in Fig. \ref{fig:circuittwport}(a) represents the model of a generic system of two inductively coupled resonators. Assuming that the resonant frequencies of both resonators are equal to $\omega_0$, $Z_{21}=ix_{21}$. Neglecting the source resistance $R_s$, which is usually small, the $\textnormal{FOM}$ reduces to $x_{21}/\sqrt{r_{11}r_{22}}=\omega_0M/\sqrt{R_1R_2}$. Hence to improve the FOM, one strives to maximize the value of $x_n$, which can be achieved via, (1) the increase of the mutual coupling (moving the resonators closer, better design of the coils to increase the flux linkage, or the use of metamaterials or near field plates to focus the magnetic field inside the receiver coil\cite{yerazunis,huang2012jap,urzhumovprb,grbicnfp,ranaweerajap}.) and (2) reduction of the self resistors $r_{ii}$ or, in another words, the increase of the resonators $Q$ factors. It is worth noting that the absence of the $r_n$ is intrinsic to the scheme shown in Fig. \ref{fig:circuittwport}(a). If a relay resonator is inserted between the transmitter and receiver as shown in Fig. \ref{fig:circuittwport}(b), $Z_{21}$ becomes real at resonance. Therefore, the FOM is enhanced by simultaneously increasing the numerator and reducing the denomenator of (\ref{eq:kQnorm}). \added{It is worth to note that in the midfield regime, where the separation distance between source and load is comparable to the wavelength, power is transferred via inductive and radiative modes, resulting in a complex $Z_{21}$. The optimum frequency of WPT schemes in human tissues was obtained through the maximization of a figure of merit parameter that mainly depends on $|Z_{21}|$\cite{poon}.}

It is worth to briefly illustrate how optimal conditions are calculated from the two port representation. For a detailed discussion and derivations, the reader can refer to Refs. \onlinecite{dionigi2015rigorous,minnaert2017single,ohira2014maximum,ohira2014extended}. As a first step, the network is represented via its $\mathbf{Z}$ parameters, which provide a convenient means for expressing the system efficiency $\eta$ in terms of the two port parameters and load impedance. \added{Accordingly $\eta$ assumes the form}
\begin{equation}
\eta=\frac{Re.~Z_L}{Re.~Z_{in}}\abs{\frac{Z_{12}}{Z_{22}+Z_L}}^2,
\end{equation}
\added{where $Z_{in}=Z_{11}-Z_{12}^2/(Z_{22}+Z_L)$ is the input impedance. The above expression is valid for an arbitrary load impedance $Z_L$. To find the optimal efficiency for the given network ($\mathbf{Z}$ is fixed), the derivative of the efficiency w.r.t the load impedance $Z_L$ is set to zero and the optimal load is determined. In general, the optimal load is complex and takes the form}
\begin{equation}
X_L=r_{22}\theta_x-x_{22}
\end{equation}
and
\begin{equation}
R_L=r_{22}\theta_r,
\end{equation}
\added{where $r_{22}$ and $x_{22}$ are the real and imaginary of $Z_{22}$, $\theta_r=\sqrt{1+x_n^2}\sqrt{1-r_n^2}$ and $\theta_x=r_nx_n$. Additionally, it can be shown that, still under the maximum efficiency condition, the power delivered to the load can be maximized whenever the network parameters satisfy,}
\begin{equation}
\label{eq:condition1}
\frac{x_{11}}{x_{12}}=\frac{r_{12}}{r_{22}}.
\end{equation}
 An arbitrary network does not necessarily satisfy (\ref{eq:condition1}). However by inserting a series reactive element $X_c$ to $x_{11}$ or the input port (refer to Fig. \ref{fig:circuittwport} (c)), the condition (\ref{eq:condition1}) can be met. The reactive element acts like a compensator (and hence the $c$ subscript) that absorbs no power, but is necessary for maximizing the power delivered to the network and load.

\added{On the other hand, if} one is interested in maximizing the power transfer to the load rather than maximizing efficiency, similar steps can be taken to calculate the optimal load and corresponding efficiency. \added{The optimum load in this case, however, can be determined by invoking the maximum power transfer theorem to show that the optimum load is equal to the conjugate of the impedance of the two port network seen from the load terminal.} It is important to point out that the condition of maximum efficiency does not imply maximum power transferred to the load or vice versa. In fact as will be seen later in Section \ref{sec:Measurements}, for a given input available power the higher the efficiency is, the less the power transferred to the load.

The interaction between the resonators is usually studied using CMT\cite{Karalis2008,awai2005,awai2007,Elnaggar2015ECMT}. From a CMT point of view, the insertion of a relay coil between the source and receiver improves the FOM via the introduction of a non-bonding (dark) mode that enables EIT-like transmission\cite{Elnaggar2017JAP,Hamam2009}. Therefore, a circuit model that directly exposes the modal behaviour and at the same time preserves the same two port parameters is quite valuable.  Not only does it support the understanding of the system behaviour, it also serves as a unified framework that can be used to compare the performance of multi-moded systems with the convenient inductive resonant coupling schemes.
\begin{figure*}[!htb]
\center
\includegraphics[width=2\columnwidth]{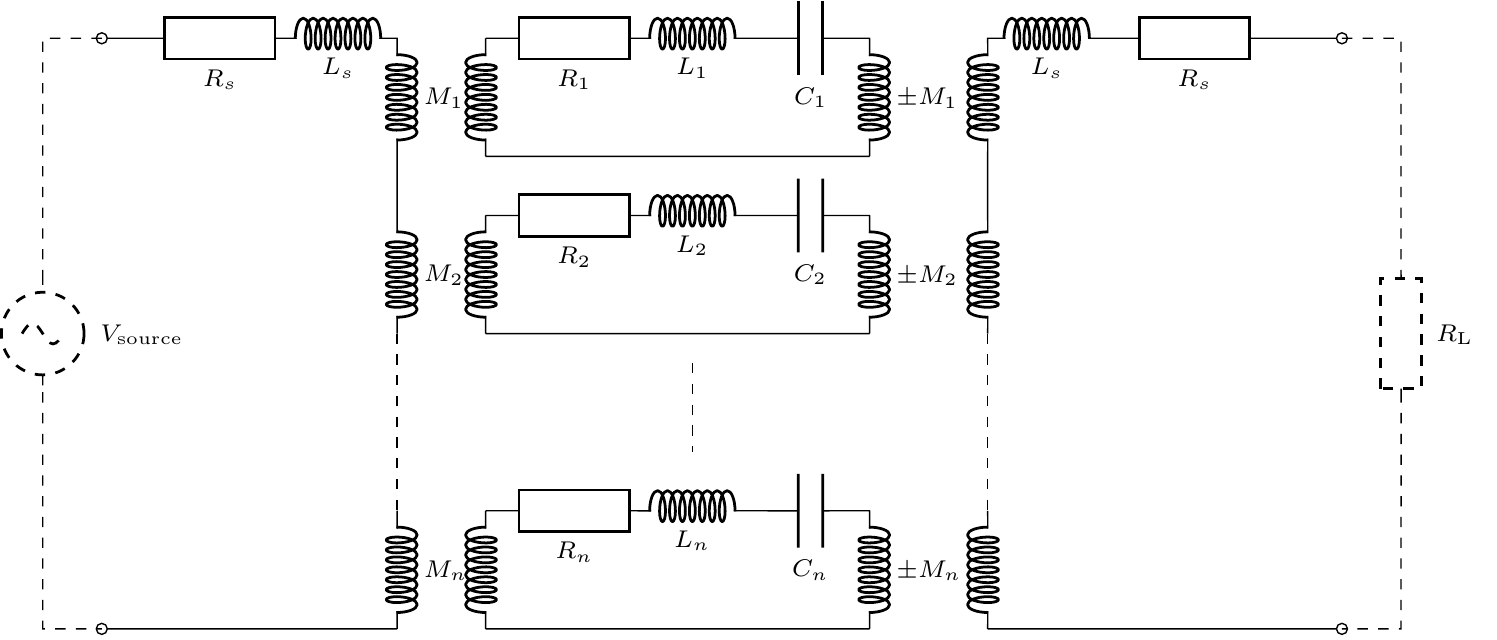}
\caption{A circuit representation of the excitation of a general symmetric resonator. $L_s$ and $R_s$ represents the equivalent lumped of the feeding inductive loop. $M_i$ is the coupling of the input power to the $i^\textnormal{th}$ mode. Depending on the mode symmetry, the mode couples with load network via an equivalent mutual coupling $\pm M_i$. }
\label{fig:ModalCCT}
\end{figure*}

Fig. \ref{fig:ModalCCT} represents the circuit model of a generic resonator\cite{CollinsFoundations}. By a \emph{resonator} here we mean a generic system that exhibits resonant modes resulting from the solution of the Helmholtz's wave equation subjected to suitable boundary conditions. For instance, it can be considered a microwave cavity or a collection of lumped LC circuits; or a combination of both. The excitation is coupled to the resonators' modes via a loop, which is modelled by a series resistor and inductor, $R_s$ and $L_s$. The mutual inductances $M_k$ represent the mutual coupling between the excitation and the $k^\textnormal{th}$ mode, which is generally determined by the overlap of the source with the mode fields profile. When a given mode is excited, its magnetic field intercepts the loop connected to the load. Each resonant mode is represented by an RLC circuit and the coupling to the load is modeled by a mutual inductance $\pm M_k$, where the sign depends on the symmetrical nature of the mode. From an input-output prespective, the circuit can be treated as a two port network where

\begin{equation}
\label{eq:Z11general}
Z_{11}=Z_{22}=R_s+i\omega L_s+\sum_{k=1}^N\frac{\omega^2M_k^2}{Z_k}
\end{equation}
and
\begin{equation}
\label{eq:Z21general}
Z_{21}=Z_{12}=\sum_{k=1}^N(-1)^{p_k}\frac{\omega^2M_k^2}{Z_k},
\end{equation}
where $Z_k\triangleq R_k+i\omega L_k-i/\omega C_k$, $N$ is the number of excited modes, and $p_k=0$ ($p_k=1$) when the mode fields have an even (odd) symmetry around the symmetry plane. Eqs. (\ref{eq:Z11general}) and (\ref{eq:Z21general}) express $Z_{11}$ and $Z_{21}$ as the net effect of all excited modes. Furthermore, the symmetry of a mode can be identified via the sign of $r_{21}$ in the vicinity of the mode resonant frequency.

\begin{figure*}[!htb]
\center
\includegraphics[width=2\columnwidth]{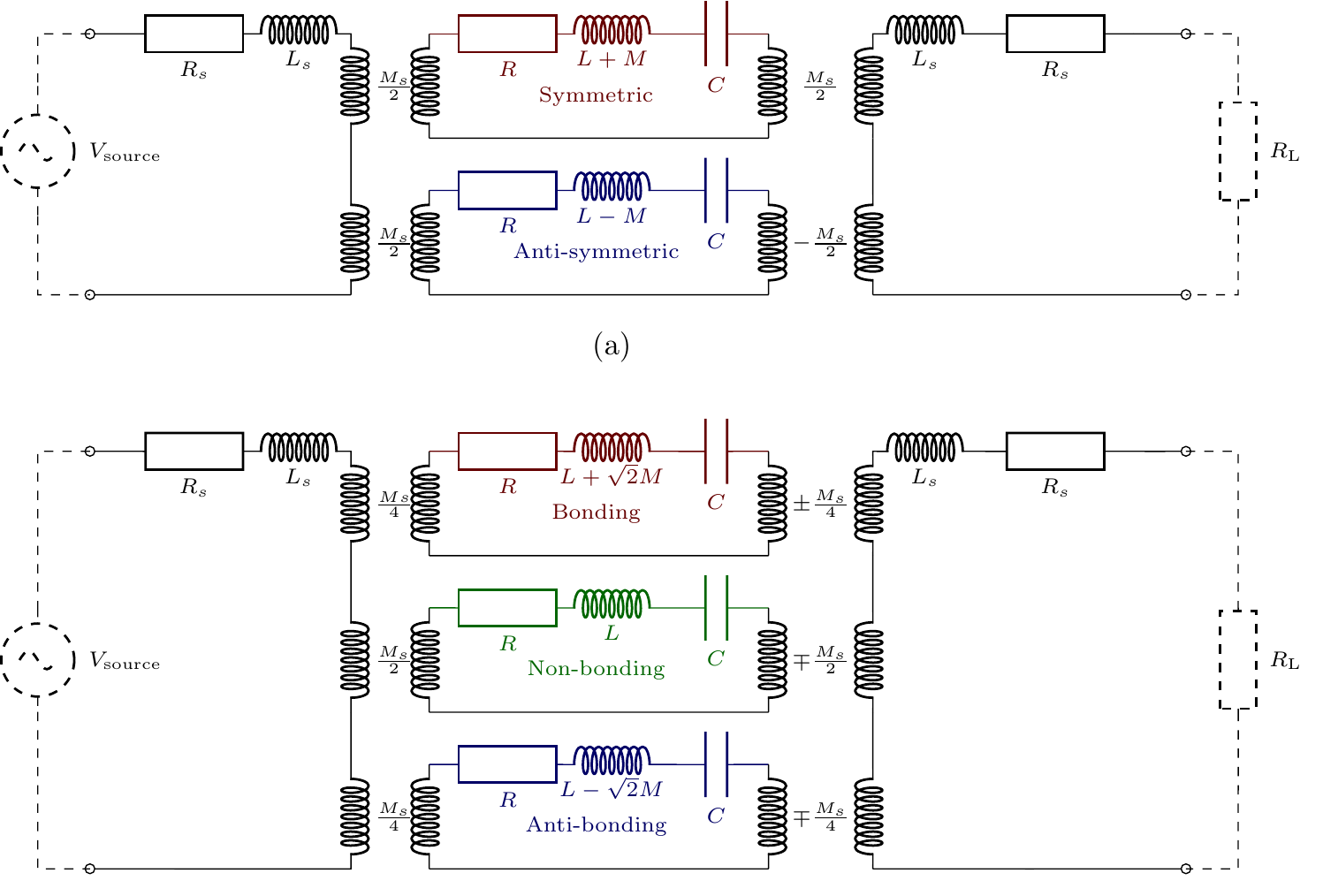}
\caption{A modal equivalent circuit of (a) the inductive resonant coupling WPT shown in Fig. \ref{fig:circuittwport}(a). (b) The three coupled resonators of Fig. \ref{fig:circuittwport}(b). }
\label{fig:ModalInductive}
\end{figure*}

It may appear that inductive resonant coupling systems that are usually represented by the circuits in Figs. \ref{fig:circuittwport}(a) and (b) are different in nature from the resonator circuit shown in Fig. \ref{fig:ModalCCT}. Nonetheless, it can be easily shown that for identical and/or high $Q$ resonators, Figs. \ref{fig:circuittwport}(a) and (b) reduce to the modal circuits presented in Fig. \ref{fig:ModalInductive}(a) and \ref{fig:ModalInductive}(b), respectively. For convenience, we assume that the two resonant coils appearing in Fig. \ref{fig:circuittwport}(a) are identical (i.e, $R_2=R_1=R, L_2=L_1=L,C_2=C_1=C$). Fig. \ref{fig:ModalInductive}(a) presents the energy transfer mechanism from the source to the load via the resonant modes: symmetric and anti-symmetric, which can be visualized as two parallel pathways that connect the source to the load. The two modes couple identically with the source ($M_1=M_2$) and oppositely with the load, due to the different symmetry of both modes. When $\omega=\omega_0=1/\sqrt{LC}$, $r_n=0$ and the $\textnormal{FOM}$ depends on $x_n$ only.

On the other hand, the insertion of a relay coil, as shown in Fig. \ref{fig:circuittwport}(b), results in three coupled modes: bonding $\left(\omega_b\triangleq 1/\sqrt{(L+\sqrt{2}M)C}\right)$, non-bonding $\left(\omega_n\triangleq 1/\sqrt{LC}\right)$; and anti-bonding $\left(\omega_a\triangleq 1/\sqrt{(L-\sqrt{2}M)C}\right)$. Interestingly, the mutual coupling of the non-bonding with the source is twice that of any of the other two modes. The non-bonding mode allows an electromagnetic induced transparency-like mechanism to transfer energy between a transmitter and receiver\cite{Hamam2009,Elnaggar2017JAP}. The bonding and anti-bonding modes are analog to the symmetric and anti-symmetric modes of a two inductively coupled resonant system.

\subsection{Identification of Modes Dynamics}
\label{subsec:ident}
Measurements are often made in the frequency domain, where the scattering parameters at the reference planes are measured by a Vector Network Analyzer. For the circuit in Fig. \ref{fig:ModalCCT}, measurements represent the scattering parameters of the full two port between the source and load (neglecting the attached connectors). The scattering parameters can be converted into any other set of network parameter. For instance, they can be converted to the $\mathbf{Z}$ parameters, which provide a convenient representation.  The measured two ports network parameters, however, represent the net effect of the structure as seen at the terminals. For the resonator modal circuit (Fig. \ref{fig:ModalCCT}), it is desirable, as has been previously mentioned, to decompose (unfold) the net effect into the modal constituents.

From a circuit modelling perspective, a resonant mode $m$ is characterized by its resonant frequency $\omega_{0m}$ and $Q$ factor $Q_m$. To fully describe the multi-moded resonator in Fig. \ref{fig:ModalCCT}, the resonant frequencies $\omega_{0m}$, $Q_m$, mutual inductances $M_m$, and modal polarities $p_m$ must be extracted from the measured data. The modal parameters can be easily extracted if only one mode is excited such that the effect of other modes is usually small. If the higher order modes cannot be ignored, their collective effect in the vicinity of the given mode is generally complex in nature. Traditionally, the effect has been modelled by a reactance $X(\omega)$ that is approximated by the first two terms of Taylor's series (i.e, $X(\omega)\approx X(\omega_0)+X_1(\omega_0)(\omega-\omega_0)$, where $X_1(\omega_0)\triangleq \partial X/\partial\omega |_{\omega=\omega_0}$). Taking $X(\omega)$ into account, the mode resonant frequency and the unloaded $Q$ can be obtained from the measured scattering parameters\cite{kajfez,sun1995unloaded}. For the muti-moded cavity case, however, the modes are very close to one another such that in the vicinity of a given mode, the effect of other nearby modes can widely vary and more terms in the Taylor's series need to be retained. Additionally for an accurate construction of the circuit representation (\ref{eq:Z11general}) and (\ref{eq:Z21general}), the identification process must be repeated for all excited modes. To overcome the challenges of the traditional identification approach, we apply a different identification procedure that relies on fitting the frequency domain $Z_{21}$ to a rational function such that, in the vicinity of a given mode frequency, it behaves much like (\ref{eq:Z21general}). It is worth noting that such approach has been previously applied to extract resonators' parameters used for microwave filtering applications\cite{liao2007vector,zhao2016model}.

The measured $Z_{21}$ frequency domain data is fitted to a rational function $\hat{Z}_{21}$ using the vector fitting (VF) method \cite{gustavsen1999rational}

\begin{equation}
\hat{Z}_{21}(s)=\sum_{k=1}^N \frac{C_ks+D_k}{s^2+\hat{\omega}_{0k}s/\hat{Q}_k+\hat{\omega}_{0k}^2},
\end{equation} 
where $\hat{\omega}_{0k}$ and $\hat{Q}_k$ are the identified resonant frequency and $Q$ factor of the $k^\textnormal{th}$ pole, respectively. \added{Each rational function in the above expression results from the combination of a simple first order rational function of the form $A_k/(s-p_k)$ with its complex conjugate $A_k^*/(s-p_k^*)$}. \deleted{As the}\added{From a fixed} $m^\textnormal{th}$ resonant mode \deleted{is concerned}\added{perspective}, $\hat{Z}_{21}$ can be re-written as
\begin{equation}
\label{eq:Z21hatatwm}
\hat{Z}_{21}(\omega)=\hat{F}(\omega)+\frac{iC_m\omega+D_m}{-\omega^2+i\hat{\omega}_{0m}\omega/\hat{Q}_m+\hat{\omega}_{0m}^2},
\end{equation}
where 
\begin{equation}
\hat{F}(\omega)\triangleq \sum_{k\neq m} \frac{iC_k\omega+D_k}{-\omega^2+i\hat{\omega}_{0k}\omega/\hat{Q}_k+\hat{\omega}_{0k}^2}.
\end{equation}
In the vicinity of $\hat{\omega}_{0m}$ (i.e, $\delta\triangleq \omega-\hat{\omega}_{0m}$ such that $|\delta|\ll \hat{\omega}_{0m}$),  both the numerator $n(\delta)$ and denomenator $d(\delta)$ of the second term in (\ref{eq:Z21hatatwm}) become first order in $\delta$. In another words, (\ref{eq:Z21hatatwm}) is approximated to 
\begin{equation}
\label{eq:Z21hatapprox}
\hat{Z}_{21}\approx \hat{F}'(\omega)+\frac{\hat{\mathcal{A}}_m}{\delta\left[2\hat{\omega}_{0m}-i\hat{\omega}_{0m}/\hat{Q}_m\right]-i\hat{\omega}_{0m}^2/\hat{Q}_m}.
\end{equation}
The exact expressions of $\hat{F}'(\omega)$ and $\hat{\mathcal{A}}_m$ are readily obtained from (\ref{eq:Z21hatatwm}) after the substitution $\omega =\hat{\omega}_{0m}+\delta$ is made and noting that $n(\delta)/d(\delta)=k+r/d(\delta)$, where $k$ is a constant. Finally, $k$ is absorbed in $\hat{F}'(\omega)$ (i.e, $\hat{F}'(\omega)=\hat{F}(\omega)+k$). When $\hat{Q}_m\gg 1$, the behaviour is dominated by the denomenator of the second term in (\ref{eq:Z21hatapprox}).

Now we show that the circuit model given by (\ref{eq:Z11general}) and (\ref{eq:Z21general}) reduces to (\ref{eq:Z21hatapprox}) at the vicinity of the $m^\textnormal{th}$ mode. To prove this, it should be observed that $Z_{21}$ can be written as

\begin{equation}
\label{eq:Z21approx}
Z_{21}(\omega)=F(\omega)-\frac{(-1)^{p_m}i\omega^3M_m^2/L_m}{\omega^2-i\omega\omega_{0m}/Q_{0m}+\omega_{0m}^2}.
\end{equation}
In the vicinity of $\omega_{0m}$ and after neglecting higher order terms of $\delta$, (\ref{eq:Z21approx}) reduces to
\begin{equation}
Z_{21}(\omega)\approx F'(\omega)+\frac{{\mathcal{A}}_m}{\delta\left[2{\omega}_{0m}-i{\omega}_{0m}/{Q}_m\right]-i{\omega}_{0m}^2/{Q}_m}.
\end{equation}
Eqs. (\ref{eq:Z21approx}) and (\ref{eq:Z21hatapprox}) take the same form. For a large $Q_m$ value and whenever the frequencies of the other modes are outside the $m^\textnormal{th}$ mode bandwidth, $F'(\omega)$ and $\hat{F}'(\omega)$ are \emph{relatively} slow varying functions of frequency. Hence, the behaviour is mainly determined by the second term (fast varying in the vicinity of $\omega_{0m}$). Therefore, whenever VF correctly estimates the value of the resonant frequency ($\omega_{0m}$), the identical forms of the expressions (\ref{eq:Z21approx}) and (\ref{eq:Z21hatapprox}) imply that the unloaded $Q$ of the $m^\textnormal{th}$ mode ($Q_m$) is $\hat{Q}_m$.

\subsection{Identification of the total response}
As discussed in Subsection \ref{subsec:ident}, the resonant frequencies and the unloaded $Q$ of the different modes are identified using VF. To determine the general behaviour, it is also essential to calculate the strength of coupling of each mode to the source ($M_k$). This can be achieved by defining, 
\begin{equation}
\psi_k(\omega)\triangleq\frac{i\omega^3}{-\omega^2+\omega\omega_{0k}/Q_k+\omega_{0k}^2}.
\end{equation}
Hence, $Z_{21}$ can be written as a linear superposition of $\psi_k$,
\begin{equation}
Z_{21}=\sum_{k=1}^Na_k\psi_k(\omega),
\end{equation}
where $a_k\triangleq \zeta M_k(-1)^{p_k}$ and $\zeta \triangleq M_k/L_k$ is a coupling coefficient term. The basis functions $\psi_k$ are complex ($\psi_k(\omega)=\textnormal{Re.}\psi_k+i\textnormal{Im.}\psi_k$) therefore, we determine the coefficients $a_k$ by expanding $r_{21}$ in the real part of $\psi_k$ and apply the least mean square fitting method. Once the coefficients $a_k$ are determined, the imaginary part of $\psi_k$ can be used to validate the accuracy of the circuit model in Fig. (\ref{fig:ModalCCT}). In another words, the $a_k$ values are determined from the fitting of the measured $r_{21}$ to $\sum_{k=1}^N a_k \textnormal{Re.~}\psi_k(\omega)$. For the model (\ref{fig:ModalCCT}) to be accurate $x_{21}$ should be equal to $\sum_{k=1}^N a_k\textnormal{Im.~}\psi_k(\omega)$.

\section{SCR and DR2SCR configurations and modes}
\label{sec:scrdr2scr}

For a complete description of the system, we identify the excited 3D EM modes via full-wave simulation. Such analysis provides a deep insight that guides the understanding of the DRs/SCR interactions; particularly when EIT-like transfer becomes observable.

\begin{figure}[!htb]
\centerline{\includegraphics[width=0.6\columnwidth]{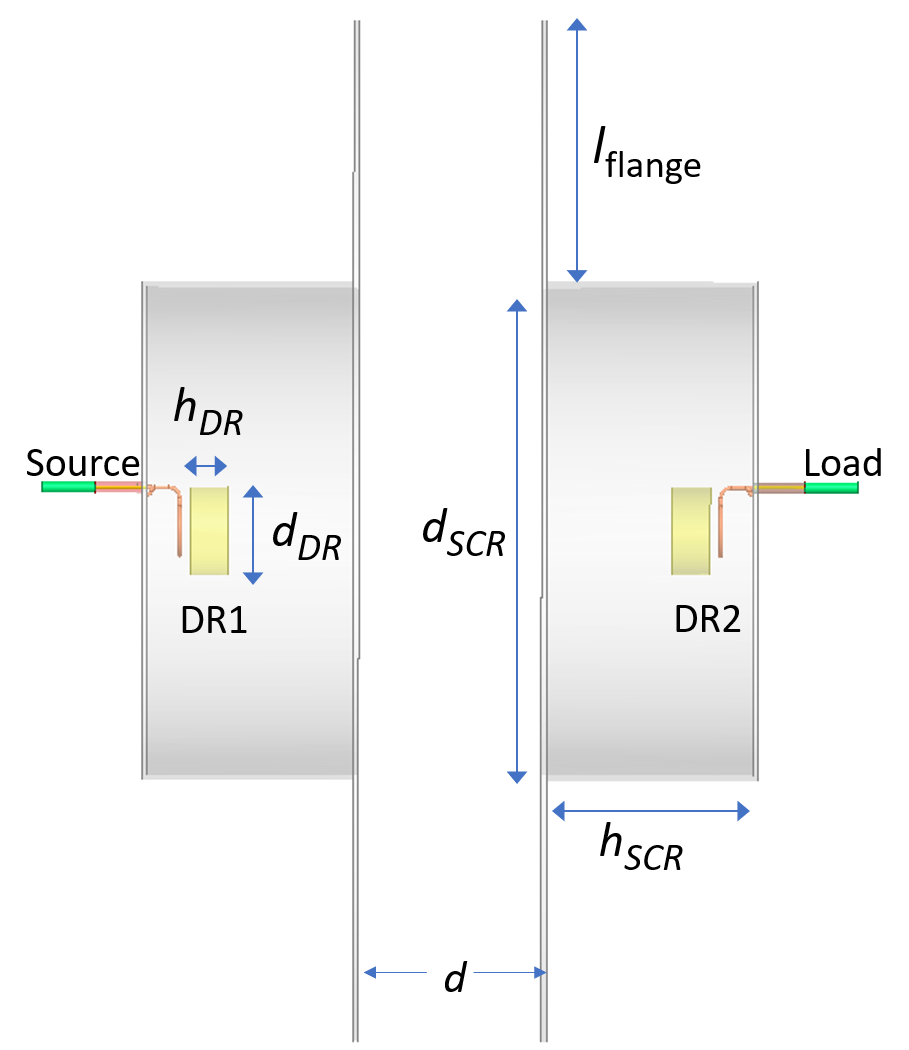}}

\caption{The Split Cavity Resonator with the dielectric resonators. Dimensions: \deleted{$d_\textnormal{SCR}=37.4~cm$}$d_\textnormal{SCR}=18.7 \textnormal{ cm}$, $h_\textnormal{SCR}= 8.2\textnormal{ cm}$, $d= 7\textnormal{ cm}$, $l_\textnormal{flange}=10 \textnormal{ cm}$, $h_\textnormal{DR}=1.4\textnormal{ cm}$ and $d_\textnormal{DR}=2.5~\textnormal{ cm}$. \added{The DRs are placed at a distance of 1.7 cm from the SCR flat surfaces.}}
\label{fig:SCR}
\end{figure}

An SCR consists of two separate conducting cylindrical halves as shown in Fig. \ref{fig:SCR}. Unlike coupled resonators, both halves need to be present to sustain the resonant modes. In other words, the modes are due to the SCR as a whole and are not considered to result from the interaction of the two separate subsystems. The exact excited modes depend on the system geometry, dimensions, the excitation and how it couples to the modes profiles. The rather complicated set of modes allows $Z_{21}$ to be complex.

In the limit where the distance between the two halves $d\rightarrow 0$, the structure forms a cylindrical cavity. Hence, the SCR modes can be visualized as \emph{perturbed} cylindrical cavity modes. The flanges extending in the lateral direction form a radial two parallel plates waveguide. As long as the distance $d$ is smaller than the cut-off frequency of the waveguide, the fields of the resonant modes are evanescent and stay localized inside the structure. The waveguide has a $TE_{10}$ dominant mode with a cut off frequency $f_\textnormal{cut}=c/2d$. \added{The purpose of the flange is to reduce the EM leakage through the gap. As long as the field profile accross the gap is below the mode cut off frequency, the radial propagation constant is imaginary (i.e, the wave is evanescent). The longer the flange, the less the fields magnitude at the flange edges and the less the radiation from the structure. This is due to the fact the fields exponentially decay in the radial direction. In Ref. \onlinecite{Elnaggar2017JAP}, it was shown that the flange length can be reduced with no significant effect on performance.} 

The source is fed via a coaxial probe from one end, where it inductively couples the input power to the SCR modes via a coupling loop. Two DRs can be inserted in the SCR. Due to the interaction of the DRs and SCR modes, efficient power transfer can be achieved as was theoretically demonstrated in Ref. \onlinecite{Elnaggar2017JAP} and shown below via simulations and measurements. The excited modes intercept the receiving coil, which pumps energy to the load connected to the output coaxial cable.

Generally speaking, at any given frequency, the fields inside the SCR (and DRs, when present) are the superposition of the excited modes. As the input frequency approaches the frequency of a particular mode (or pole), the fields of that given mode become dominant, provided it is efficiently coupled to the excitation loops and the frequencies of other modes are sufficiently far. The exact full-wave analysis can be quite complex. Fortunately for our purpose, a grey box model is sufficient for determining the system behaviour.  Furthermore, the modal circuit discussed in section \ref{sec:Theory}, along with the procedure of calculating the contribution of each mode to the terminal parameters, enable a scalar analysis of the modal behaviour.

\subsection{Empty Split Cavity Resonator}
\begin{figure}[!htb]
\centerline{\includegraphics[width=1\columnwidth]{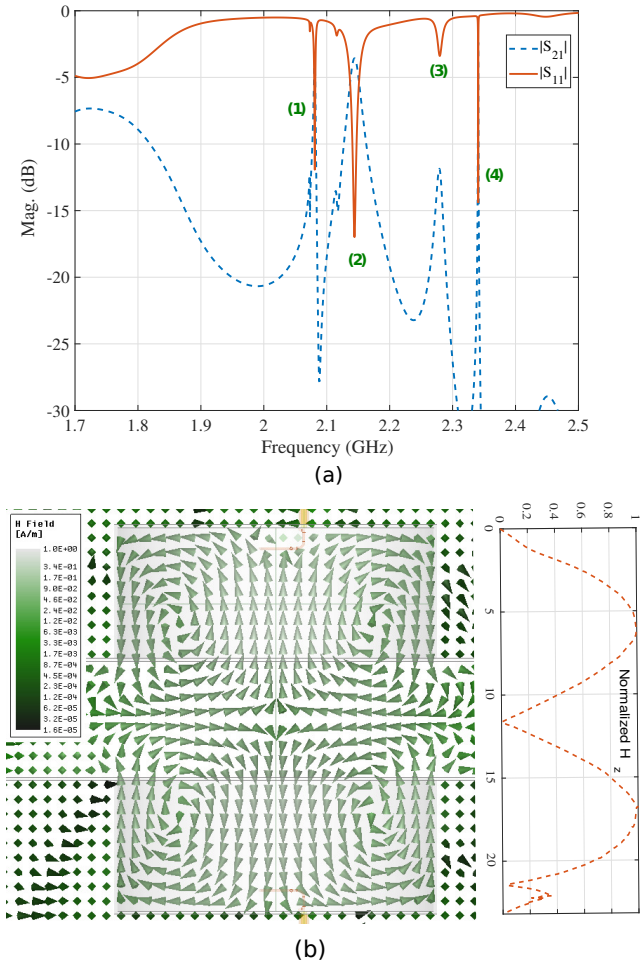}}
\caption{(a) Simulated S parameters of the SCR when displaced 7 cm a part. (1)- (4) denote the positions of the resonant modes. (1): $TM_{111}$ at $f=2.081~\textnormal{GHz}$, (2):$HEM_{111}$ at $f=2.145~\textnormal{GHz}$, (3): $TM_{112}$ at $f=2.28~\textnormal{GHz}$ and (4): $TE_{012}$ at $f=2.34~\textnormal{GHz}$. (b) Simulated magnetic field along the axial axis $H_z$ of the $TE_{012}$ mode.}
\label{fig:empyscrsim}
\end{figure}

The SCR, shown in Fig. \ref{fig:SCR}, is simulated using HFSS (Ansys\textsuperscript\textregistered, Electromagnetic Suite, 2016). The dimensions  are given in the figure. The structure is excited via an ideal coaxial cable that is connected to a current loop. The current loop enables $TE_{0mn}$ modes to be efficiently excited, however other modes can couple to the source as well. Fig. \ref{fig:empyscrsim} (a) presents the magnitude of the simulated $S_{11}$ and $S_{21}$ parameters. The dips in the $S_{11}$ represent the excited modes as highlighted in the figure. Inspecting the field profile of  the second mode reveals that it is an HEM mode. The fourth mode is a $TE_{012}$ mode with a frequency $f_{TE012}=2.34~\textnormal{GHz}$. This mode has a very high $Q$; in fact, the eigenmode solution predicts its $Q\sim 10,000$. Due to its extremely narrow bandwidth, exciting this mode may be challenging. Nevertheless, the $TE_{012}$ mode strongly couples with DRs $TE_{01\delta}$ modes, as will be shown in the next subsection. In the axial direction, the fields have two nulls (hence the mode is described by two half wavelengths in the axial direction, Fig. \ref{fig:empyscrsim}(b)). When $d=7~\textnormal{cm}$, $f_\textnormal{cut}=~2.143~\textnormal{GHz}$. Although $f_\textnormal{TE012}>f_\textnormal{cut}$, the fields of the $TE_{012}$ do not propagate through the radial waveguide since its field profile does not match the   waveguide $TE_{10}$ mode profile.  \added{However if the flange is excessively short, the resonant modes will have lower $Q$ values due to the added radiative losses. To be more specific, the $TE_{20}$ is the relevant waveguide mode, which has a cut-off frequency of $f_{cut}=c/d\approx 4.3 \textnormal{ GHz}$. Therefore, the mode is evancesent with an attenuation constant of $k_r=\sqrt{k_z^2-k_0^2}=75.7~ \textnormal{m}^{-1}$. For a 10 cm flange (the one used here), the field decays by more than 65 dB from its value at the SCR inner radius. Reducing the flange length to 5 cm results in a decay of approx. 32 dB. Therefore, the flange length can be reduced to 5 cm (or even 3 cm) with no significant effect on performance. (For a parametric analysis on the effects of the flange length, disturbances due to presence of dielectric objects, and change in dimension, the reader may refer to Refs. \onlinecite{wptc} and \onlinecite{imarc}.)}

\added{The separation distance $d$ affects the frequency and $Q$ factor of the SCR resonant modes. The larger $d$ is, the lower the resonant frequency and $Q$ are (due to the increased radiation from the gap). When $d$ is slightly perturbed around its nominal value, the circuit parameters and $\eta$ will slightly change. However as the change in $d$ becomes larger, the modes constituents may change. For instance, for smaller $d$ values, some higher order modes will be pushed further up in frequency and hence gets \emph{barely} excited. On the other hand, for larger $d$ values, the frequencies of higher order modes reduce and may become close to the excitation frequency. For extremely large values of $d$ (i.e, $d\gg d_{cut-off}$), the SCR modes will become radiative with very low $Q$ values. This means that the different modes will overlap and modal picture will be blurred. At the vicinity of the DRs $TE_{01\delta}$ modes, the non-bonding mode permits the system to tolerate changes in the dimensions. This is because the non-bonding mode does not depend on the SCR modes. (For an elaborate discussion on the permissible magnitude of detuning, please refer to Ref. \onlinecite{wptc}).}

\deleted{By inspecting the $S_{11}$ and $S_{21}$, the transfer efficiency, when the termination resistance is $50~\Omega$, can be calculated as}
\added{In general the input power of a general two port network is given by}
\begin{equation}
P_{in}=|a_1|^2-|b_1|^2,
\end{equation}
\added{where $a_1$ and $b_1$ are the amplitude of the incident and reflected waves, respectively. The output power is similarly given by}
\begin{equation}
P_{out}=|b_2|^2-|a_2|^2,
\end{equation}
\added{for a matched load (i.e, terminated with a $50~\Omega$ impedance), $a_2=0$ and $P_{in}$ becomes $|a_1|^2\left(1-|S_{11}|^2\right)$. Therefore, the efficiency $\eta$ for this particular load becomes,}
\begin{equation}
\label{eq:eta50}
\eta_{50\Omega}\triangleq\frac{P_{out}}{P_{in}}=\frac{|b_2|^2}{|a_1|^2\left(1-|S_{11}|^2\right)}=\frac{|S_{21}|^2}{1-|S_{11}|^2},
\end{equation}
\added{where by definition $S_{21}\triangleq b_2/a_1$ and $S_{11}\triangleq b_1/a_1$, when $a_2=0$.}

It is important to note, however, that the efficiency given by (\ref{eq:eta50}) is not the maximum efficiency of the system; it is merely the efficiency when the load is fixed to $50~\Omega$.

\subsection{SCR with Dielectric Resonators}
\label{subsec:DRs}

\begin{figure}[!htb]
\centerline{\includegraphics[width=\columnwidth]{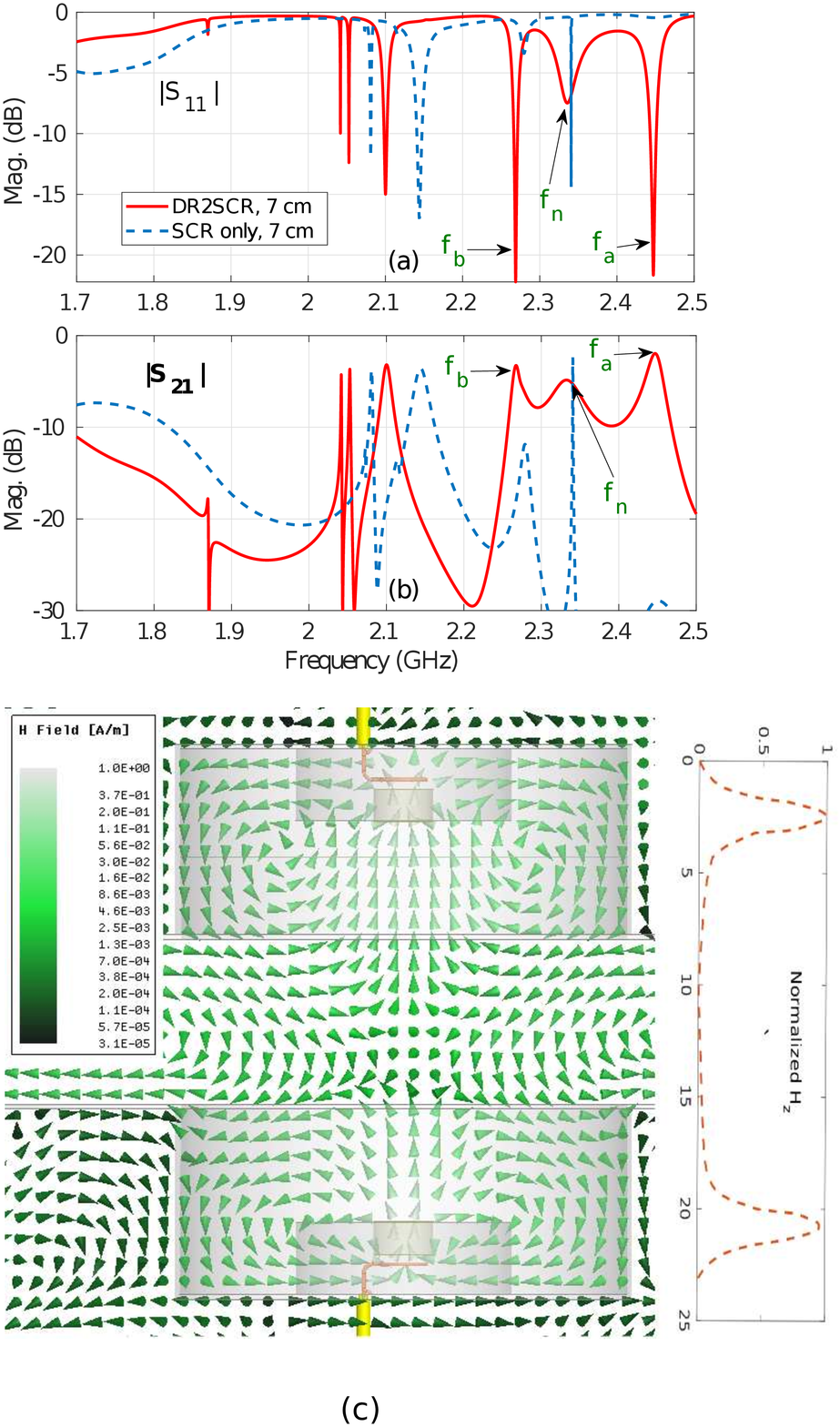}}
\caption{(a), (b) Simulated S parameters of the DR2SCR when displaced 7 cm a part and compared to the simulated S parameters of the SCR only. $f_b= 2.27~\textnormal{GHz}$ is the bonding mode, $f_n= 2.336~\textnormal{GHz}$ non bonding mode and $f_a=2.448~\textnormal{GHz}$ is the anti-bonding mode resulting from the interaction between the DRs $TE_{01\delta}$ modes and the SCR $TE_{012}$ mode. (c) Simulated magnetic field of the non-bonding mode inside the SCR and along the axis of symmetry. \added{The grey boxes surrounding the DRs represent low loss low dielectric constant foams that hold the DRs in place.}}
\label{fig:dr2scrsim}
\end{figure}

Two DRs of $\epsilon_r=25$ and $\tan\delta=0.002$ ($Q=500$) are inserted in the SCR. They are supported by a low loss/low $\epsilon_r$ foam and placed close to the coupling loops to enable the excitation of their $TE_{01\delta}$ modes. The DRs dimensions were chosen such that the $TE_{01\delta}$ mode has a resonance frequency of $\approx 2.3~\textnormal{GHz}$ and, therefore, can strongly couple with the SCR $TE_{012}$ mode. Fig. \ref{fig:dr2scrsim} shows the simulated results compared to the case when the SCR was empty. Due to the high $\epsilon_r$ value, the SCR frequencies between $2-2.2~\textnormal{GHz}$ are shifted down. At the vicinity of $2.3~\textnormal{GHz}$, three modes are clearly visible in the spectrum. The SCR $TE_{012}$ and two DRs $TE_{01\delta}$ interact, resulting in three coupled modes\cite{Elnaggar2017JAP,AMR2017}. Very close to the $TE_{012}$ frequency, appears the non-bonding mode, which does not have an SCR component. The $Q$ of the non-bonding mode is lower than that of the $TE_{012}$ mode, resulting in a wider bandwidth in $|S_{11}|$ and $|S_{21}|$. The bonding (anti-bonding) mode is formed at a frequency lower (higher) than the non-bonding mode.

The coupling between the modes results in a significant change in the fields profile. Fig. \ref{fig:dr2scrsim}(c) shows the magnetic field profile at the non-bonding frequency. The magnetic field $H_z$ along the axial direction verifies that the fields are solely due to the DR modes with no sinusoidal components coming from the SCR modes. Similar behaviour was previously observed and studied using CMT, when the DRs interacted with the SCR $TE_{011}$ mode\cite{Elnaggar2017JAP}.
\section{Measurements}
\label{sec:Measurements}
\added{The system shown in Fig. \ref{fig:SCR} is fabricated from a low profile material. The DRs are cut from a 1 inch cylindrical rod of $\epsilon_r=25\pm 10\%$ and $\tan\delta<0.002$ (Eccostock\textsuperscript{\textregistered} HIK500F). The DRs are supported in place via a low $\epsilon_r$ foam, not shown in the Figure. Coupling to energy source and load is provided via miniature loops, centred with the DRs to excite the $TE_{01\delta}$ modes.}

\begin{figure}[!htb]
\centerline{\includegraphics[width=\columnwidth]{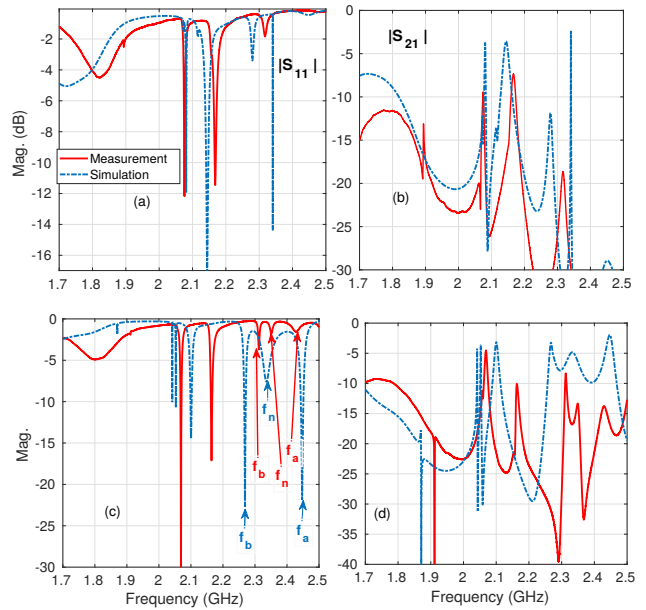}}
\caption{Measurement vs Simulation results when the two SCR halves are placed 7 cm apart. SCR (a) and (b) $S_{11}$ and $S_{21}$, respectively. DR2SCR (c) and (d) $S_{11}$ and $S_{21}$, respectively. The exact positions of the bonding, non-bonding and anti-bonding modes are highlighted in (c).}
\label{fig:simvsmeas}
\end{figure}

The measured $S_{11}$ and $S_{21}$, with the separation distance $d=7\textnormal{ cm}$, are reported in Fig. \ref{fig:simvsmeas}. The dips in $S_{11}$ represent the excited modes. There is a good correlation between measurement and simulation. Discrepancies between measurements and simulations are attributed to the following. Firstly, a realistic model of the SMA connectors was not taken into account. The mismatch and delay of the connectors introduce a slight shift in the scattering parameters. Secondly, measurements show that $S_{11}$ and $S_{22}$ are slightly different due to the tolerances in fabricating and aligning the two SCR halves. Furthermore, the tolerance in $\epsilon_r$ of the DRs adds to the uncertainity, where the $TE_{01\delta}$ frequency can be any value between 2.2 - 2.42 GHz; with a nominal value of 2.3 GHz. The exact value of $\epsilon_r$ also affects the strength of coupling with the SCR $TE_{012}$ mode. This in turn is reflected in how $f_b$ and $f_a$ are positioned with respect to $f_n$.

For the SCR measurement, the dip of the $TE_{012}$ was missed from $S_{11}$ during the VNA frequency sweep cycle. This is due to the very high $Q$ value of the mode; emphasizing the difficulty of relying on the mode for WPT, regardless of its high $Q$ value. However, more desirable effects can be harnessed when the DRs are included; as shown in Figs. \ref{fig:simvsmeas}(c) and (d). Indeed, in the DR2SCR configuration, the interaction of the of the DRs with the SCR are clearly observed in the measured spectra. The interaction was previously studied when the DR modes were tuned to the SCR $TE_{011}$ mode. In the current situation, the interaction is more complex. As a first order description, it can be approximated by the coupling of three modes. However, the presence of other modes inevitably contributes to the net response. Additionally, the mismatch between the two SCR halves will result in asymmetric modes, unlike the ones depicted in Figs. \ref{fig:empyscrsim} and \ref{fig:dr2scrsim}.

\begin{figure*}[!htb]
\centerline{\includegraphics[width=1.6\columnwidth]{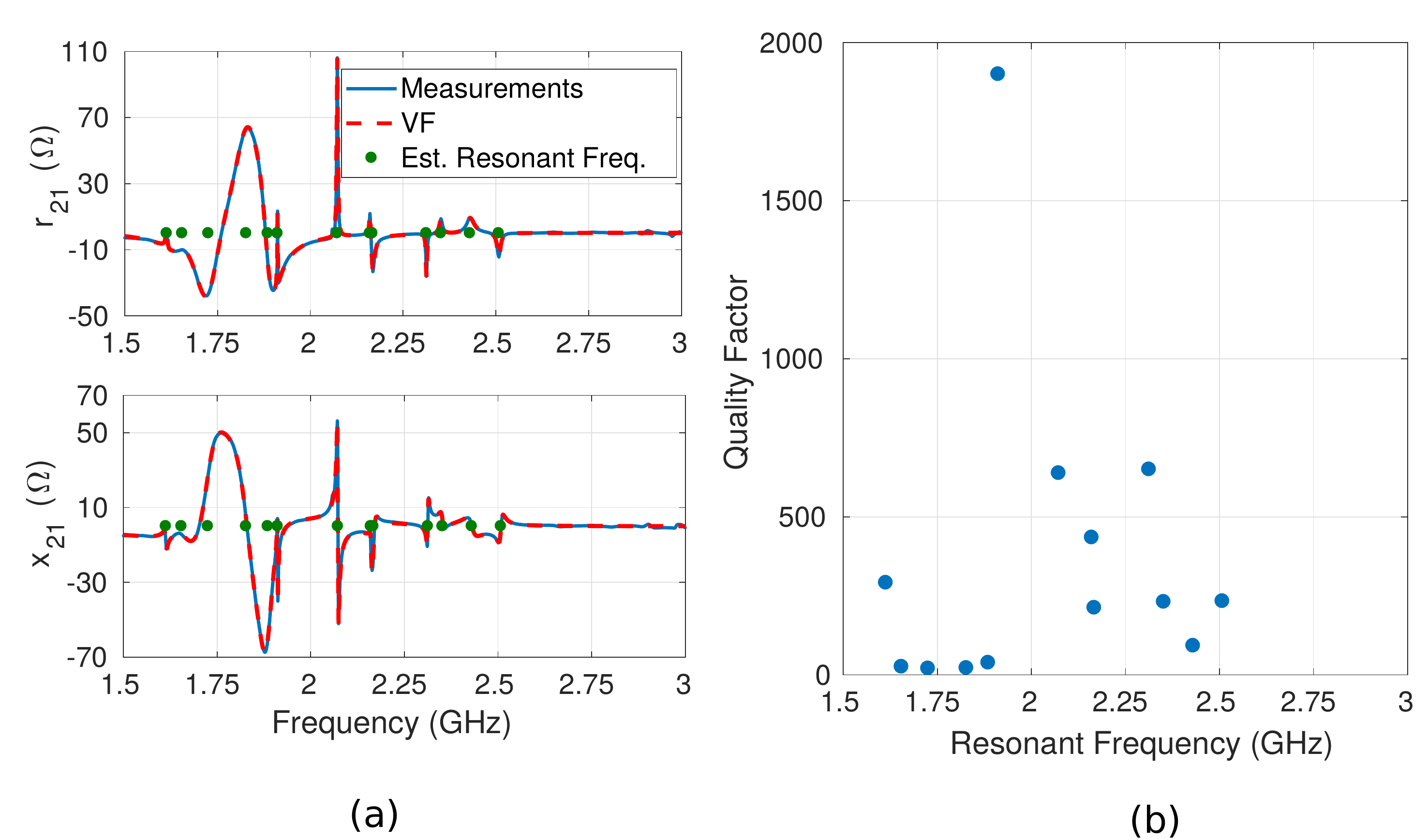}}
\caption{(a) The real (top) and imaginary (bottom) components of the measured $Z_{21}$ fitted to a rational function using VF. The imaginary of the poles (resonant frequencies $\hat{\omega}_{0k}$) are also shown. (b) The Quality factor of the poles ($\hat{Q}_m$).}
\label{fig:VF}
\end{figure*}

\begin{figure}[!htb]
\centerline{\includegraphics[width=0.9\columnwidth]{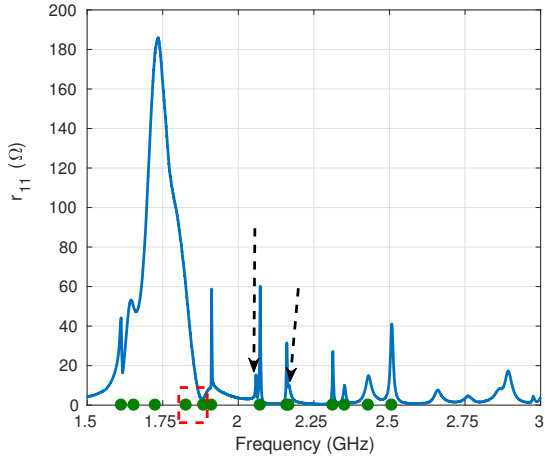}}
\caption{The measured $r_{11}$ vs frequency with the fitted poles highlighted.}
\label{fig:r11vspoles}
\end{figure}

To overcome the above challenges, we adopt the use of the modal circuit introduced in Section \ref{sec:Theory}. \added{Hereafter, if not explicitly mentioned, the focus will be on the DR2SCR system.} Vector Fitting is used to fit $Z_{21}$ to a rational function as described in Section \ref{sec:Theory}. The results are plotted in Fig. \ref{fig:VF}(a). The maximum error between the measured data and the rational function does not exceed -30 dB. The imaginary parts of the poles determined by VF are highlighted. Generally speaking, there is no guarantee that the poles correspond to the system natural frequencies. However, whenever they do, the estimated $Q$s correspond to the system $Q$s as was shown in Section \ref{sec:Theory}. To verify that the estimated poles do indeed correspond to the natural frequencies, we use the fact that for high $Q$ modes, $r_{11}$ attains its maxima at or sufficiently close to the natural frequencies. Fig. \ref{fig:r11vspoles} presents the measured $r_{11}$, where the estimated natural frequencies $\hat{\omega}_{0k}$ are shown on the abscissa. It is clear that out of 13 complex poles, two of them (highlighted by the dashed box) appear not to represent physical system poles. Fortunately, this will have a local impact only. As long as mode $m$ is concerned, the effect of other modes is encapsulated in the slow varying function $F(\omega)$ and VF guarantees an accurate representation of it, regardless of the fact that some poles may have not been correctly estimated. Furthermore for frequencies below 2.5 GHz, two modes were not detected, as highlighted by the two arrows in Fig. \ref{fig:r11vspoles}. It is clear that the two modes have a narrow bandwidth and are very close to the frequencies of other detected modes and it can safely be assumed that their effect will be absorbed in the calculation of the parameters of the nearby modes. All other poles correspond to natural modes. Hence, $Q_k=\hat{Q}_k$. Fig. \ref{fig:VF}(b) presents $\hat{Q}_k$. At the low frequency end, $\hat{Q}_k$ are very small, therefore according to the discussion in Section \ref{sec:Theory} the estimated parameters may not be very accuate in this frequency regime. \added{It is worth to mention here that the estimated $Q$ of the mode at around 1.9 GHz is significantly high. As shown in Appendix B, this mode corresponds to an anti-resonance, which needs to be represented by a shunt resonator. Fortunately at the frequency range around 2.3 GHz, the unaccurately estimated $\hat{Q}_k$ values and the anti-resonance are considered local effects that are absorbed in the $\hat{F}'(\omega)$ term in (\ref{eq:Z21hatapprox}).}

\begin{figure*}[!htb]
\centerline{\includegraphics[width=1.6\columnwidth]{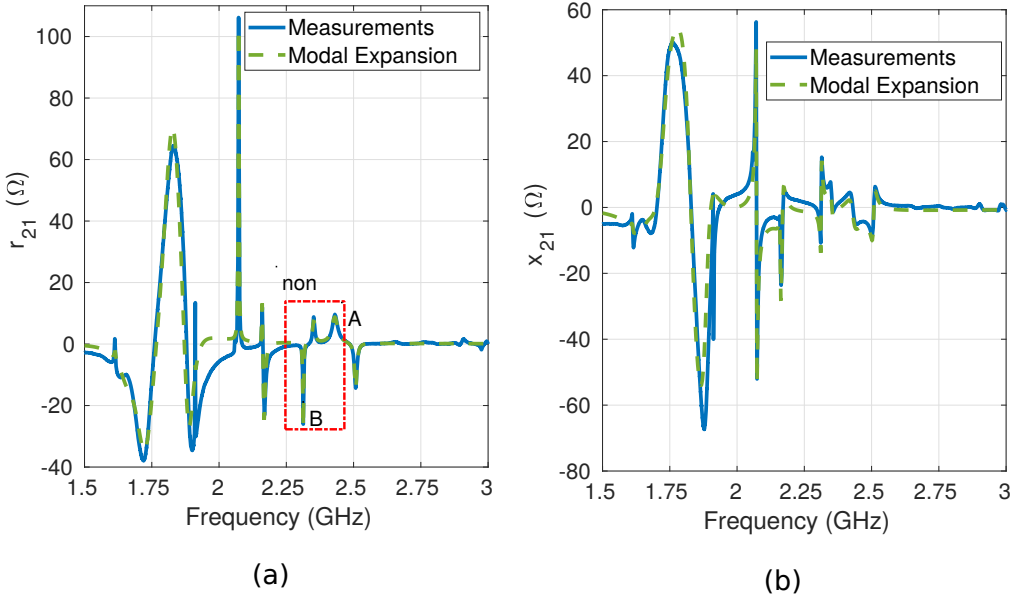}}
\caption{Measured $Z_{21}$ versus the modal decomposition model given by (\ref{eq:Z21general}). (a) Real of $Z_{21}$, $r_{21}$ and (b) Imaginary of $Z_{21}$, $x_{21}$.}
\label{fig:r21x21modal}
\end{figure*}

Once the modal $\omega_k$ and $Q_k$ values are estimated, least mean square fitting is then used to find the complete $r_{21}$ over the whole frequency range. The behaviour is depicted in Fig. \ref{fig:r21x21modal}(a). The ability to fit the measured data to (\ref{eq:Z21general}) over a relatively wide bandwidth (1.5 - 3 GHz) implies that the circuit model in Fig. \ref{fig:ModalCCT} accurately captures the system dynamics. Additionally, the calculated $a_k$ parameters were used to compute $x_{21}$ from the imaginary part of $\psi_k$ as shown in Fig. \ref{fig:r21x21modal}(b). The excellent fit between the circuit model and the measured data strengthen the belief in the validity of the schematic in Fig. \ref{fig:ModalCCT}. In this case, there are $N=13$ different modes that couple the source to load.

It is worth noting that although the circuit model relies on the terminal characteristics, the modal decomposition of equations (\ref{eq:Z11general}) and (\ref{eq:Z21general}) allow the symmetrical nature of the modes to be determined as exhibited by the sign of $r_{21}$. Furthermore, the different RLC resonators represent the coupled modes resulting from the interaction of the DRs with the SCR modes. Noting that the DRs $TE_{01\delta}$ and SCR $TE_{012}$ modes have approximately the same resonant frequencies, their interaction can be approximated by the three coupled system in Fig. \ref{fig:circuittwport}(b) and its modal equivalent (Fig. \ref{fig:ModalInductive}(b)). Due to the odd symmetry of the $TE_{012}$ it can be shown using CMT (Appendix) that the non-bonding mode has an even symmetry, or equivalently $r_{21}>0$. From Fig. \ref{fig:ModalInductive}(b), it is clear that at the anti-bonding (bonding) frequency, $r_{21}>0$ ($r_{21}<0$) (the signs of the mutual couplings of the non-bonding and the anti-bonding are the same, opposite to that of the bonding mode). The response of $r_{21}$ in the vicinity of the $TE_{012}$ mode is highlighted in Fig. \ref{fig:r21x21modal}(a) by the dashed box.

\begin{figure}[!htb]
\centerline{\includegraphics[width=\columnwidth]{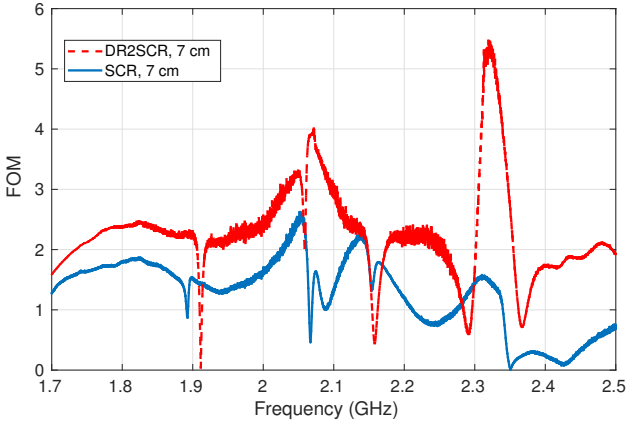}}
\caption{FOM=$\sqrt{({r_n^2+x_n^2})/({1-r_n^2})}$ calculated from the measured $\mathbf{Z}$ parameters for the SCR only and DR2SCR, when $d= 7\textnormal{ cm}$.}
\label{fig:FOM}
\end{figure}

\begin{figure}[!htb]
\centerline{\includegraphics[width=\columnwidth]{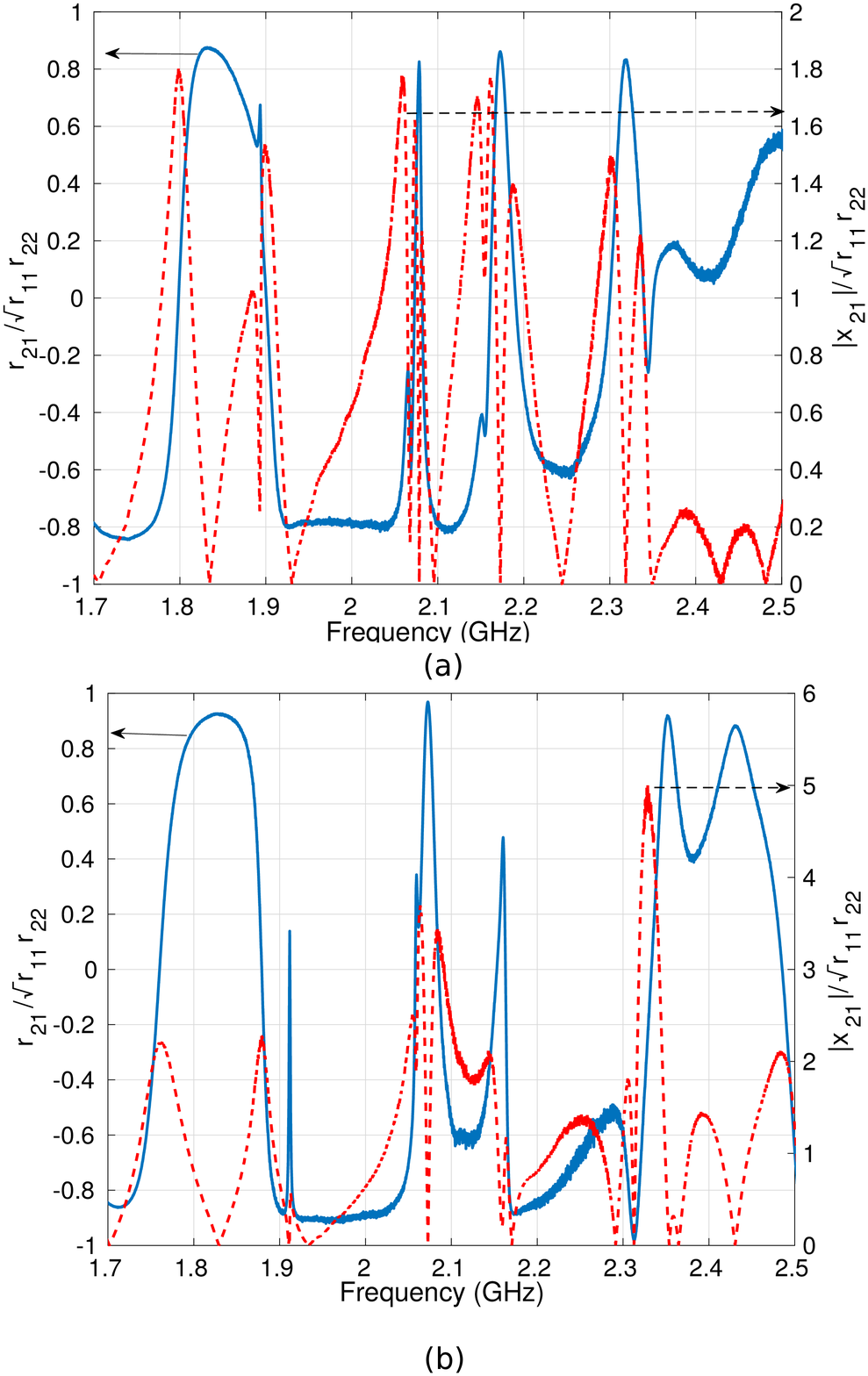}}
\caption{$r_n=r_{21}/\sqrt{r_{11}r_{22}}$ and $x_n=x_{21}/\sqrt{r_{11}r_{22}}$ calculated from measurements. (a) SCR, 7 cm. (b) DR2SCR, 7 cm.}
\label{fig:rnxn}
\end{figure}

The FOM is calculated from the $\mathbf{Z}$ parameters using (\ref{eq:kQnorm}) as shown in Fig. \ref{fig:FOM}. As Fig. \ref{fig:FOM} clearly shows, the FOM is significantly enhanced when the DRs are inserted. As expected, it attains its maximum value in the vicinity of the $TE_{012}$ due to the interaction of the $TE_{012}$ mode with the DRs $TE_{01\delta}$ modes. \added{When the DRs are inserted, the FOM at 1.9 GHz is basically zero, emphasizing the fact the mode is anti-resonance.} Moreover, Fig. \ref{fig:rnxn} shows the contribution of $r_n$ and $x_n$ to the net FOM.

 It is interesting to point out that for the SCR only case, the magnitude of $x_n$ is roughly double that of $r_n$. Both $r_n$ and $x_n$ change rapidly near 2.3- 2.4 GHz, possibly peaking at the $TE_{012}$ mode. For the DR2SCR case, the situation becomes more interesting. The magnitude of $x_n$ is roughly increased by three folds due to the modal interaction between the DRs and the SCR, which is reactive in nature. Additionally, $r_n$ drops to values very close to -1 and then starts to go up. This makes the denominator of (\ref{eq:kQnorm}) very small and hence reinforce the increase in FOM. For this particular configuration, however, the $x_n$ approaches zero as $r_n$ approaches -1. Nevertheless, the net effect of $r_n$ and $x_n$ results in a peak of FOM in the vicinity of the $TE_{012}$ mode. The two degrees of freedom offered by the possibility of controlling both $r_n$ and $x_n$, provide insight on how to enhance the performance of WPT systems; this is to be compared to inductive resonant coupling, where only one tunable parameter  $x_n$ is possible. The modal decomposition of $Z_{21}$ and $Z_{11}$ shows that the behaviour of FOM in the vicinity of some mode $m$ is affected mainly by the $m^\textnormal{th}$ mode and to a lesser extent by the other modes as reflected by their resonant frequencies, $Q$ and the mutual coupling to both the source and load. Therefore, by the proper \emph{placement} of the resonators' modes, it is potentially possible to maximize the FOM. This is equivalent to \emph{engineering} the parallel pathways in Fig. \ref{fig:ModalCCT} in a way that allows the EM fields to constructively add at the load.  It is important to note, however, that $x_n$ and $r_n$ are not fully independent and strongly depend on the physical realizability of the network. For instance, $x_{21}$ and $r_{21}$ are always related by the Kramers-Kroning relations \cite{wing2008classical}; hence adding constraints that need to be taken into account during the design process.

\begin{figure}[!htb]
\centerline{\includegraphics[width=\columnwidth]{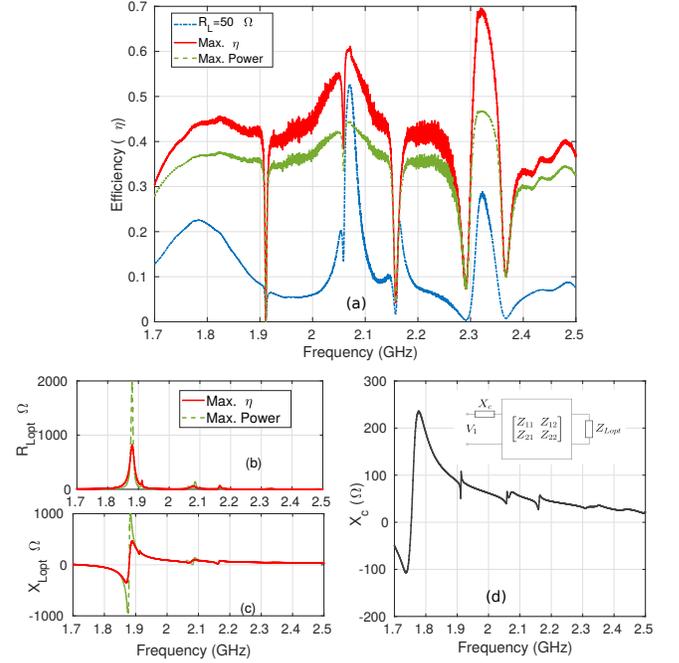}}
\caption{DR2SCR, 7 cm a part. (a) Maximum efficiency, efficiency at max. load power and efficiency determined directly from VNA measurement (i.e, efficiency when load impedance $Z_L= 50~\Omega$. (b) Real and Imaginary load values $R_\textnormal{Lopt}$ and $X_\textnormal{Lopt}$, respectively at maximum efficiency and max load power. (c) Value of series compensator reactance $X_c=$. The inset illustrates where the compensator element is added to the two port network input terminal.}
\label{fig:efficiency1}
\end{figure}

The efficiency is calculated for three different cases: (1) $50~\Omega$ termination- the efficiency- calculated directly from the VNA measurements, (2) the maximum efficiency and (3) the efficiency when maximum power is absorbed by the load. As Fig. \ref{fig:efficiency1} (a) shows, the efficiency can wildly change over the frequency range of interest for the three cases. The maximum efficiency goes to around 0.7 in the vicinity of the $TE_{012}$ mode. As was already discussed at the frequency range, the modes behaviours are mainly due to the interaction between the DRs and SCR $TE_{012}$ modes. Therefore, it is expected that if a lower $\tan\delta$ DRs are used, the efficiency can further be improved. \added{It is worth to note that at the anti-resonance mode at 1.9 GHz, the efficiency is practically zero.} Additionally, significant losses are expected to be attributed to the low profile connector and the foam supporting the DRs. Fig. \ref{fig:efficiency1} (b) demonstrates that the optimal load impedance for maximum efficiency and maximum power transfer are generally not the same, in agreement with the theoretical predictions \cite{dionigi2015rigorous}. The compensator reactive element required at 2.3 GHz is found to be $\sim 30~\Omega$ or equivalently an inductor of inductance $\sim 2-3 \textnormal{ nH}$.

\begin{figure}[!htb]
\centerline{\includegraphics[width=\columnwidth]{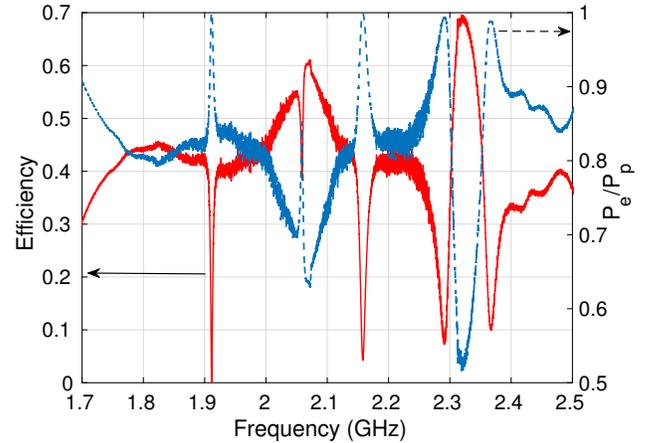}}
\caption{Maximum efficiency and power at maximum efficiency to maximum power at load ratio.}
\label{fig:efficiency2}
\end{figure}

As was previously shown based on the theoretical analysis of two port networks, the efficiency at maximum load power is less than the maximum efficiency. Hence, to deliver more power to the load (i.e, operating the source near its rating value), the efficiency drops. Conversely at maximum efficiency, the actual power absorbed by the load is a fraction of the maximum power that can be delivered. Fig. \ref{fig:efficiency2} illustrates this inverse relation. For instance at around $2.3 \textnormal{ GHz}$, the power at maximum efficiency $P_e$ to the maximum power that can potentially be delivered to the load $P_p$ is $\approx 0.05$.

\begin{figure}[!htb]
\centerline{\includegraphics[width=\columnwidth]{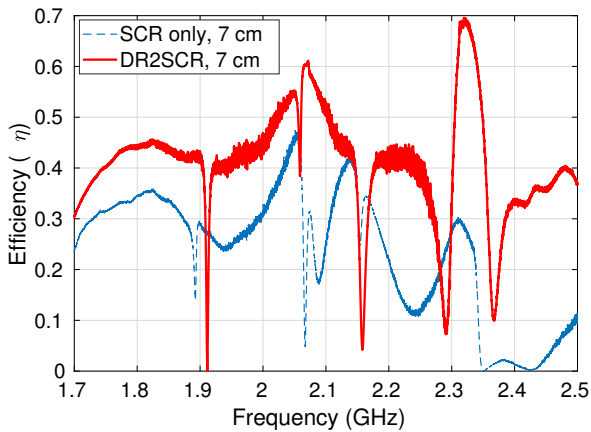}}
\caption{Maximum attainable efficiency for SCR and DR2SCR when the separation distance $d= 7\textnormal{ cm}$.}
\label{fig:eta_scr_dr2scr}
\end{figure}

Finally, Fig. \ref{fig:eta_scr_dr2scr} show the maximum efficiency of the DR2SCR compared to the SCR. Regardless of the DRs added losses due to their $\tan\delta$, the DRs have improved the performance, particularly when they strongly interact with the SCR $TE_{012}$ mode as was previously discussed. From a circuit theory point of view, the DR2SCR configuration allows a simultaneous change of both the real and imaginary parts of $Z_{21}$, which in turn improves the FOM.

\section{Conclusion}
A Wireless Power Transfer system based on the interaction of DRs and a multi-moded SCR has been proposed. Due to the complexity of the structure, a circuit based approach has been employed to study the system performance. It is shown that such configuration, unlike conventional inductive resonant coupling, offers the possibility of controlling both the real and imaginary parts of the transfer impedance $Z_{21}$. This in turn has resulted in a significant improvement of the system figure of merit and efficiency. Furthermore, a modal circuit description is used for the first time to decompose the terminal parameters to their modal constituents. Hence, it reduces the task of manipulating $Z_{21}$ to the engineering of the different modes.

\added{The multi-moded SCR explored provides a platform to examine the interaction of different modes and how they contribute to the terminal transfer impedance $Z_{21}$. It may find applications where conducting surfaces naturally appear. One potential application is charging electric vehicles via the exploitation of the already existing conducting surfaces. The dimension of the SCR may be reduced, for instance, by coating the surfaces with a dielectric material or via the manipulation of the boundary conditions using artificial materials (metamaterials) such as artificial magnetic conductors\cite{sievenpiper}. Additionally, the modal profile and field distribution may be controlled via the manipulation of the effective permittivity and permeability; for instance by using epsilon and/or mu near zero materials\cite{enghetaprl,lipworth}.}

\section*{Appendix A - Non-bonding mode}
The non-bonding mode resulting from the coupling between the two DRs and the SCR modes is briefly presented here. Generally speaking, the fields of the coupled modes can be expressed as the linear combination of the DRs and SCR modes. Close to the $f_{TE012}\approx f_{TE01\delta}$, the interaction is dominated by the DRs $TE_{01\delta}$ and SCR $TE_{012}$ modes. Hence around $f_{TE012}$, the field profile is mainly determined by these three modes. The DR two modes will be labeled "1" and "3"; and the SCR $TE_{012}$ is denoted by the "2" subscript. Therefore the analysis is identical to that in Ref. \onlinecite{Elnaggar2017JAP}. It is worth noting, however, that unlike Ref. \onlinecite{Elnaggar2017JAP}, the interaction is anti-symmetric (i.e, $\kappa_{12}=-\kappa_{32}=\kappa$). Taking this into account the eigenvalue problem is written as,
\begin{equation}
\begin{bmatrix}
f_0^2 & -f_0^2\kappa &0\\
-f_0^2\kappa & f_0^2 &f_0^2\kappa\\
0 & f_0^2\kappa &f_0^2
\end{bmatrix}
\begin{bmatrix}
a_1\\
a_2\\
a_3
\end{bmatrix}=
f^2
\begin{bmatrix}
a_1\\
a_2\\
a_3
\end{bmatrix},
\end{equation}
where $f_0$ is the resonant frequency of the $TE_{012}$ mode, $a_i$ is the expansion coefficient of the $i^\textnormal{th}$ mode, and $f$ is the to be determined coupled frequency. There are three eigenmodes for the above equation: bonding, non-bonding and anti-bonding. By inspection, it is readily found that $f=f_0$ satisfies the eigenvalue problem with an eigenvector
\begin{equation}
\mathbf{a}_\textnormal{n}=
\frac{1}{\sqrt{2}}\begin{bmatrix}
1\\
0\\
1
\end{bmatrix}, 
\end{equation}
which is the eigenvector of the non-bonding mode. The fields in both DRs are in phase as shown in Fig. \ref{fig:dr2scrsim}.

\section*{Appendix B - Nature of mode at the frequency of the SCR $TE_{011}$ mode}
The sharp dip of the magnitude of $S_{21}$ at the frequency of the SCR $TE_{011}$ mode (around 1.9 GHz in Fig. \ref{fig:simvsmeas}(d)) acts as a notch filter that blocks transmission. Interestingly, the dip appears only when the DRs are inserted, suggesting that it results from the coupling of the DRs with the SCR $TE_{011}$ mode. As was previously shown, the coupling between two different modes is due to the interaction of the sources of one mode with the fields of the other, which is equal to the net overlap of the fields\cite{jap2015}. When the modes frequencies are different, the interaction is smaller. However as long as the net overlap is still significant, one mode leaves its fingerprint on the other in terms of remnant fields that appear in the given mode\cite{elnaggarjmr2014}. In our case, the SCR $TE_{011}$ and DRs $TE_{01\delta}$ have very different resonant frequencies, however they still interact due to the non-diminishing value of $\kappa$, leaving a small $TE_{01\delta}$ component on the $TE_{011}$ mode. We will show below that this has the effect of a dramatic reducion in the transfer efficiency.

Here we apply CMT to describe the effect of the DRs on the SCR $TE_{011}$ mode. To fix ideas, the DR closest to the excitation, the SCR, and the other DR will be denoted resonators 1,2, and 3 respectively. For a sinusoidal excitation with frequency $\omega_{in}$, the amplitude vector of the modes $\mathbf{a}=[a_1,a_2,a_3]^t$ can be determined from the solution of  \cite{samehforcedecmt,Elnaggar2017JAP}
\begin{equation}
\label{eq:cmt}
{\mathbf{a}}(\omega_{in})=i\omega_{in}\mathbf{\Phi}(\omega_{in})\mathbf{J},
\end{equation}
where $\mathbf{J}$ is a $3\times 1$ column vector that is a function of the excitation and $\mathbf{\Phi}$ is the coupling matrix \cite{Elnaggar2015ECMT}. The excitation is coupled to resonators 1 and 2. However because resonator 1 is very close to the exciting loop, it will be assumed that the excitation is mainly coupled to it (i.e, $\mathbf{J}=[J_1,0,0]^t$). This permits the finding of simple expressions as was carried out in Ref. \onlinecite{Elnaggar2017JAP}. Another justification of the assumption is that the calculation of the eigenmode reveals that the amplitude of the SCR mode is approximately 25 times larger than the amplitude of the DR modes. Therefore, if the excitation couples energy to the SCR mode, a very minute fraction of the energy will be passed to the DR closer to the load (resonator 3). The transfer efficiency depends on the ratios $|a_2/a_1|^2$ and $|a_3/a_1|^2$. Hence, $\eta$ can be written as
\begin{equation}
\label{eq:a3a1}
\eta=\frac{\sigma_L|a_3/a_1|^2}{(\sigma_L+\sigma_0)|a_3/a_1|^2+\sigma_0+\sigma_2|a_2/a_1|^2},
\end{equation}
where $\sigma_L$, $\sigma_2$ and $\sigma_0$ are the load, SCR and DR decay constants, respectively. Generally speaking, the decay constant $\sigma_i$ is equal to $\omega_i/2Q_i$. The ratios $a_2/a_1$ and $a_3/a_1$ can be found from (\ref{eq:cmt}) to be
\begin{equation}
\label{eq:a2a1}
\frac{a_2}{a_1}=\frac{(\gamma_0/\gamma_2)^2\kappa\left[1-\left(\omega_{in}/\gamma_3\right)^2\right]}{\left[1-\left(\omega_{in}/\gamma_3\right)^2\right]\left[1-\left(\omega_{in}/\gamma_2\right)^2\right]-\kappa^2}
\end{equation}
and
\begin{equation}
\frac{a_3}{a_1}=\frac{(\gamma_0/\gamma_3)^2\kappa^2}{\left[1-\left(\omega_{in}/\gamma_3\right)^2\right]\left[1-\left(\omega_{in}/\gamma_2\right)^2\right]-\kappa^2},
\end{equation}
where $\gamma_k=\omega_k+i\sigma_k$ is the complex eigen-frequency of the $k^{th}$ mode. Albeit the factor $\gamma_0/\gamma_k$, these relations are identical to the ones derived in Ref. \onlinecite{Elnaggar2017JAP}. Here, however, the resonant frequencies of the DR and SCR modes are different, which does not warrant $\gamma_0/\gamma_2=1$ approximation to be valid. For the system under consideration, $f_{TE011}\approx 1.9$ GHz, $f_{TE01\delta}\approx 2.3 $ GHz, the $TE_{011}$ $Q$, $Q_2\sim 100~-~1000$, and the loaded $Q$ of resonator 3 is $Q_3\approx 100~-~500$. To estimate the order of $\kappa$, the expression derived in Ref. \onlinecite{jmrquality} for a closed cavity will be used. Although the expression is only valid when the distance $d$ is zero, the field profile of the $TE_{011}$ does not change much; justfying its use for an order of magnitude analysis. Moreover, one needs to account for the axial shift of the DRs, where the fields of the $TE_{011}$ vary as a sinusoidal function. Taken all these into consideration, $\kappa\approx 0.02$. Plugging the above estimates into (\ref{eq:a3a1}) and (\ref{eq:a2a1}) shows that the efficiency is always less than or equal 2\%.

To consider the plausible situation where the excitation is coupled to both resonators 1 and 2, the above arguments can be modified by taking the interaction with resonator 2 into account, i.e, letting $\mathbf{J}=[J_1,J_2,0]^t$. This means that the amplitudes $a_k$ will be the linear combination of both interactions ($a_k=\Phi_{k1}J_1+\Phi_{k2}J_2$). For values of $J_2/J_1$ from 0 up to 10, it was found that still $\eta\leq 2\%$.

The anti-resonance behaviour at $f_{TE011}$ implies that the mode cannot be described via the use of the series resonance circuits in Fig. \ref{fig:ModalCCT}. For such mode, a shunt resonance circuit should be used instead.

%
\bibliography{WPTC_measurement}
\smallskip

\end{document}